\documentclass[final,3p,times]{elsarticle}
 \biboptions{comma,sort&compress}
\usepackage{here}
\usepackage{graphicx}
\usepackage{epsfig}
\usepackage{multicol}

\def\nin{\noindent}
\def\nnb{\nonumber}

\def\mysqrt{\sqrt{\frac{\omega}{\omega'}}}
\def\luv{{\Lambda_{\rm UV}}}
\def\muw{\mu_{\omega'}}

\journal{Nuc. Phys. (Proc. Suppl.)}

\begin{document}

\begin{frontmatter}



\title{The $B$-meson Light-Cone Distribution Amplitude in
  Dual Space}

$\left. \right.$\vspace{-17mm}
\begin{flushright}
{SI-HEP-2014-22, QFET-2014-16}
\end{flushright}


\author{Bj\"orn O. Lange}

\address{Theoretische Elementarteilchenphysik, Universit\"at Siegen,
  57068 Siegen, Germany. \\
{\tt lange@tp1.physik.uni-siegen.de}}

\begin{abstract}
\noindent
We describe a representation of the leading $B$-meson light-cone distribution
amplitude in heavy-quark effective theory based on the
eigenfunctions of its anomalous dimension kernel. In this
representation (called the dual LCDA) different dual momenta no longer
mix under renormalization. We discuss the perturbative and
non-perturbative nature of different regions in this space.
\end{abstract}


\end{frontmatter}


\begin{multicols}{2}
\section{Introduction}
\nin Theoretical descriptions of $B$-meson decays into exclusive light
final states often invoke a non-local matrix element in heavy-quark
effective theory (HQET) between the $B$ state and the vacuum, see
e.~g.~\cite{Beneke:1999br}. Since the light final states recoil with
considerable energy against the $B$ meson at rest (at least in some
regions of the phase space), the non-locality of the operator is
light-like. The most important of such objects is called the leading
light-cone distribution amplitude (LCDA), defined as
\cite{Grozin:1996pq}
\begin{equation}
  \label{eq:1}
  \tilde f_B m_B \;\tilde\phi_B^+(t) = \langle 0 | \bar q({ t n}) [tn,0] n\hspace{-4.8pt}/ \gamma_5 h({0})|B\rangle \;,
\end{equation}
where $\tilde f_B$ is the $B$-meson decay constant in HQET, and the
vector $n^\mu$ is light-like. (There also exists another LCDA within
the 2-particle Fock-state description of the $B$ meson, called
$\phi_B^-$, which can be discussed similarly.) The Fourier transform
of this function,
\begin{equation}
  \label{eq:2}
  \phi_B^+(\omega) = \int \frac{dt}{2\pi} e^{i\omega t}
  \tilde\phi_B^+(t) \;,
\end{equation}
is commonly used as $\omega$ represents the $n$-projection of the
light-quark's momentum. This function is principally a
non-perturbative input that enters factorization theorems and thus not
calculable in perturbation theory. However, it is possible to
calculate both its moments over a large enough intervall using an
operator product expansion (OPE) \cite{Lee:2005gza}, as well as its
dependence on the renormalization scale $\mu$, which has been
suppressed in the notation so far. Let us first consider the latter
point: the $\mu$-dependence is gouverned by the integro-differential
renormalization-group equation (RGE) \cite{Lange:2003ff}
\begin{eqnarray}
  \label{eq:2}
  \frac{d}{d\ln\mu}\phi_B^+(\omega) &=& 
  - \left[ \Gamma_c\,\ln \frac{\mu}{\omega} + \gamma_+ \right]
  \phi_B^+(\omega) \nnb \\
  && - \omega \int\limits_0^\infty d\eta \,\Gamma(\omega,\eta) \,\phi_B^+(\eta) \;.
\end{eqnarray}
The second term in this equation mixes different regions in $\omega$
when evolving in $\mu$. Therefore the solution to this equation
requires us to integrate over the full $\omega$ region of the LCDA at
the initial scale. However, as was discouvered recently the above
operation possesses a continuous set of eigenfunctions
\cite{Bell:2013tfa}
\begin{equation}
f_{\omega'}(\omega) = \sqrt{\frac{\omega}{\omega'}} J_1\left(2
  \sqrt{\frac{\omega}{\omega'}}\right)\;, 
\end{equation} 
with the parameter $\omega'$ of mass dimension 1, called the dual
momentum. The analogon for light mesons are the Gegenbauer polynomials
\cite{Efremov:1979qk,Lepage:1979zb}. A suitable representation of the
LCDA is therefore given by a linear combination of the eigenfunctions,
\begin{equation}
  \label{eq:3}
  \phi_B^+(\omega) = \int_0^\infty \frac{d\omega'}{\omega'}
  \rho_B^+(\omega') f_{\omega'}(\omega) \;,
\end{equation}
where $\rho_B^+(\omega')$ is now analogous to the Gegenbauer
coefficients in the comparison with light mesons above. The relation
(\ref{eq:3}) can be inverted using the orthogonality relations of
Besselfunctions and reads
\begin{equation}
  \label{eq:4}
  \rho_B^+(\omega') = \int_0^\infty \frac{d\omega}{\omega}
  \mysqrt  J_1 \left( 2\mysqrt \right) \phi_B^+(\omega) \;.
\end{equation}
As a result the so-defined dual\footnote{The integral transformation
  (\ref{eq:4}) of the common model $\phi_B^+(\omega) =
  \frac{\omega}{\omega_0^2} e^{-\omega/\omega_0}$ results in
  $\rho_B^+(\omega') = \frac{1}{\omega'} e^{-\omega_0/\omega'}$, which
  features a ``dual'' behaviour: whereas $\phi_B^+(\omega)$ falls off
  exponentially for large $\omega$ and vanishes linearly at the
  origin, $\rho_B^+(\omega')$ vanishes exponentially for small
  $\omega'$ and as a first inverse power for large $\omega'$.} LCDA
renormalizes locally,
\begin{equation}
  \label{eq:5}
  \frac{d}{d\ln\mu}\rho_B^+(\omega') = - \left[ \Gamma_c \ln
    \frac{\mu}{\hat \omega'} + \gamma_+ \right] \rho_B^+(\omega')\;,
\end{equation}
with the rather simple solution
\begin{equation}
  \label{eq:6}
  \rho_B^+(\omega',\mu) = e^{V(\mu,\mu_0)} \left( \frac{\mu_0}{\hat
      \omega'}\right)^{-g(\mu,\mu_0)} \rho_B^+(\omega',\mu_0) \;.
\end{equation}
Here the hatted quantity $\hat\omega' = e^{-2\gamma_E} \omega'$
denotes a rescaled dual momentum, and the functions $g$ and $V$
involve integrals over the anomalous dimensions $\Gamma_c$ and
$\gamma_+$, see e.~g.~\cite{Bell:2013tfa} for details.  Qualitatively
the functions $\phi_B^+$ and $\rho_B^+$ contain the same information;
constraints on one of them translate to constraints on the other.

\section{Perturbative constraints and large dual momenta}
\nin It is well known that the LCDA $\phi_B^+(\omega)$ does not fall
rapidly enough to have a norm. In other words, the bare matrix element in
(\ref{eq:1}) requires an extra subtraction in the local limit. Since
moments over an infinite interval are therefore not defined, one must
introduce an ultra-violet cutoff $\luv$, on which moments depend
logarithmically. Such moments, 
\begin{equation} 
  \label{eq:7}
  M_n(\luv,\mu) = \int\limits_0^\luv d\omega \,\omega^n
  \phi_B^+(\omega,\mu) \;,
\end{equation}
have been calculated in an operator product expansion in
$1/\luv$ to first order corrections as \cite{Lee:2005gza}
\end{multicols} 
{\underline{\hspace{150mm}}}
\begin{eqnarray} 
  \label{eq:momentconstraint}
    M_0 &=& 1 + \frac{\alpha_s(\mu)\, C_F}{4\pi} \left( -2 \ln^2\frac{\luv}{\mu} +2 \ln\frac{\luv}{\mu} - \frac{\pi^2}{12} \right) 
+ \frac{16\bar\Lambda}{3\luv} \frac{\alpha_s(\mu)\, C_F}{4\pi}
\left(\ln\frac{\luv}{\mu} - 1 \right) \;, \nnb \\
    M_1 &=& \luv \frac{\alpha_s(\mu)\, C_F}{4\pi} \left(-4 \ln\frac{\luv}{\mu} +6 \right)
+ \frac{4\bar\Lambda}{3} \left[ 1+ \frac{\alpha_s(\mu)\, C_F}{4\pi} \left( -2 \ln^2\frac{\luv}{\mu} +8 \ln\frac{\luv}{\mu} - \frac74 - \frac{\pi^2}{12} \right)  \right]\;.
  \end{eqnarray}
{\underline{\hspace{150mm}}}
\begin{multicols}{2}
When expressing the above in dual space we find a weighted integral
over all dual momenta
\begin{equation}
  \label{eq:8}
 M_n(\luv,\mu) = \int\limits_0^\infty d\omega'\, N_n(\omega',\luv) \,\rho_B^+(\omega',\mu)  \;,
\end{equation}
where the first few weight functions are
\begin{eqnarray}
  \label{eq:9}
  N_0 &=& \frac{\luv}{\omega'} J_2 \left(2
    \sqrt{\frac{\luv}{\omega'}}\right) \;,\\
  N_1 &=& \frac{\luv^2}{3\omega'} \left[ 2 J_2 \left(2
    \sqrt{\frac{\luv}{\omega'}}\right) - J_4 \left(2
    \sqrt{\frac{\luv}{\omega'}}\right)\right] \;.\nnb
\end{eqnarray}
They primarilly probe the dual LCDA in the region $\omega' \sim \luv$, and since
$\luv \sim \mu$ we can infer the functional dependence of $\rho_B^+$
in the region $\omega' \sim \mu$ as an expansion in $1/\omega'$ modulo
logarithms. We find up to second-order power corrections
\begin{eqnarray}
  \label{eq:10}
  \rho_B^+(\omega',\mu) &=& C_0(\ln \frac{\mu}{\hat\omega'})
  \frac{1}{\omega'} - \frac23 C_1(\ln \frac{\mu}{\hat\omega'})
  \frac{\bar\Lambda}{(\omega')^2} \;, \nnb \\ 
   C_0(L) & =& 1 + \frac{\alpha_s C_F}{4\pi} \left( -2 L^2 + 2L
     -2-\frac{\pi^2}{12} \right) \;, \nnb \\
   C_1(L) & =& 1 + \frac{\alpha_s C_F}{4\pi} \left( -2 L^2 + 2L
     +\frac54 -\frac{\pi^2}{12} \right) \;. \nnb
\end{eqnarray}

\section{Resummation of the tail}
\nin So far we note that the large $\omega'$ behaviour
of $\rho_B^+$ is $1/\omega'$. This finding and the solution
(\ref{eq:6}) seem incompatible and leads us to contemplate the
following dilemma: suppose two model builders, $A$ and $B$, are given
the task to create a model for $\rho_B^+$ at two different scales,
$\mu_A$ and $mu_B$, respectively, with $\mu_A < \mu_B$. Both feature
an asymptotic $1/\omega'$ tail. But according to (\ref{eq:6}) the tail
of the first model, $\rho_A$, will pick up a softening contribution as we evolve
from one scale to the next, and scales like $\rho_A(\mu_B) \sim
(\omega')^{-1+g(\mu_B,\mu_A)}$ at the scale of the second builder. Are
both models therefore incompatible?

The answer lies in the fact that we are comparing two different
regions in $\omega'$. Whereas $\rho_A$ has a $(\omega')^{-1}$ behaviour
around $\omega' \sim \mu_A$ and indeed evolves to a softer
$(\omega')^{-1+g(\mu_B,\mu_A)}$ dependence in this $\omega' \sim
\mu_A$ regime, we then compare it to the $\omega' \sim
\mu_B$ regime. \\

Let us therefore use standard resummation techniques by introducing an
auxiliary scale $\muw$ that scales like $\omega'$ for large
$\omega'\gg \mu$ and does not become small as $\omega'$ becomes small,
for example $\muw = \sqrt{\mu^2 + {\omega'}^2}$. 
The latter aspect has no relevance to our discussion at hand, but was
chosen so that we can avoid the Landau pole in $\alpha_s(\muw)$ when
discussing the low $\omega'$ regime later on. From (\ref{eq:6}) it
follows that
\begin{equation}
  \label{eq:11}
  \rho_B^+(\omega',\mu) = e^{-V(\muw,\mu)} \left(
    \frac{\mu}{\hat\omega'} \right)^{g(\muw,\mu)} \rho_B^+(\omega',\muw)\;.
\end{equation}
This equation allows us to state that the dual LCDA is perturbatively
calculable in resummed perturbation theory for the entire region
$\omega' \gtrsim \mu$ \cite{Feldmann:2014ika}. To see how the above
puzzle is resolved it helps to consider the function
\begin{equation}
  \label{eq:12}
  f(\omega',\mu) = \frac{d\ln \rho_B^+(\omega',\mu)}{d\ln \omega'} \;.
\end{equation}
The essence of this definition is that if $\rho_B^+ \sim
(\omega')^{-c}$ as $\omega' \to \infty$ (with $c$ a constant), then
$f\to -c$. We find
\begin{equation}
  \label{eq:13}
  f(\omega',\mu) = -1 -g(\muw,\mu) + \Gamma_c(\muw) L + \gamma_+(\muw)
  + r\;,
\end{equation}
where $L = \ln (\muw/\hat\omega')$ and $r$ collects terms of order
$\beta_0 \alpha_s^2$ and power corrections. Therefore $f \approx -1
- g(\muw,\mu)$ asymptotically for large $\omega'$ as shown in Figure~1. 
\\

\nin  \includegraphics[width=80mm]{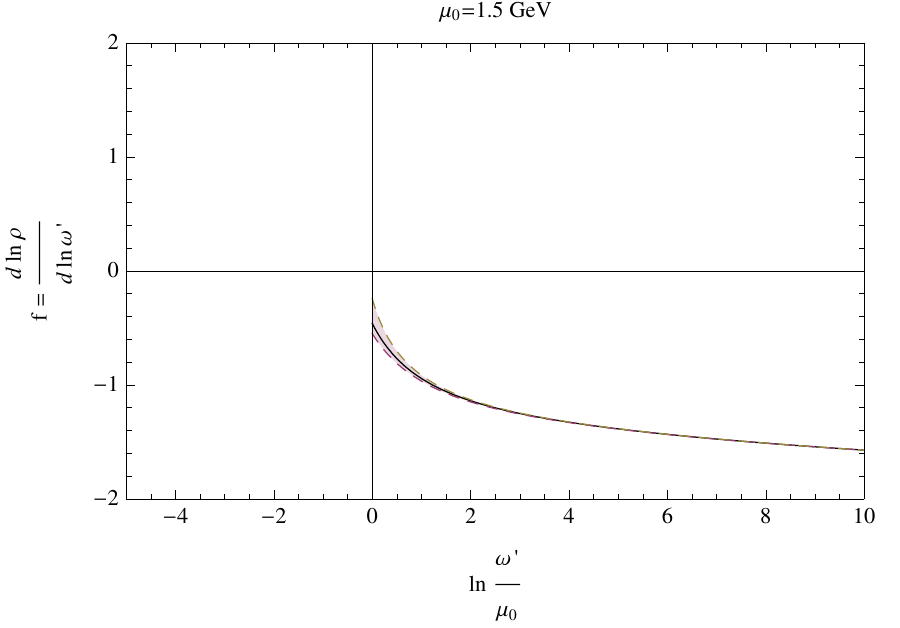}
Figure 1: Depiction of the function $f(\omega',\mu_0)$ at $\mu_0 =
1.5$ GeV and $\omega' \ge \mu_0$. \\

Since $d f(\omega',\mu) / d\ln\mu = \Gamma_c(\mu)$ (which follows from
equation (\ref{eq:5}) exactly) integrates to
\begin{equation}
  \label{eq:14}
  f(\omega',\mu_B) = f(\omega',\mu_A) + g(\mu_B,\mu_A) \;,
\end{equation}
we see that $f \approx -1$ or $\rho_B^+(\omega',\mu_B) \sim 1/\omega'$
for $\omega' \sim \mu_B$, irrespective of how large $\mu_B$ becomes. 

The finding that $\rho_B^+$ falls towards zero with an ever increasing
rate since $g(\muw,\mu)$ is a monotoneously growing function in
$\omega'$ leads to a new insight, to wit that the phenomenologically
important first inverse moment of the LCDA, $\lambda_B(\mu)$ see below, exists
at all scales $\mu$. (In fact, we can even state that any positive
moment of $\rho_B^+$ exists, since $g(\muw,\mu)$ grows with no boundary.)

Previous analyses of the LCDA \cite{Lange:2003ff,Lee:2005gza,
  Kawamura:2001jm} stated solutions for the RG evolution that broke
down as the spread between initial scale $\mu_A$ and final scale
$\mu_B$ becomes so large that $g=g(\mu_B,\mu_A) = 1$, as manifest in
factors involving $\Gamma(1-g)$ and similar functions. We now
understand that these perceived thresholds are not physical and
disappear once a resummation for large values of $\omega'$ is
included.


\section{Non-perturbative aspects and first inverse moments}
\nin In the previous narrative we have discussed the function
$\rho_B^+(\omega',\mu)$ for values of $\omega' \gtrsim \mu$, which is
the perturbative regime. For small values of $\omega'$, however,
non-perturbative physics are dominant. In the absence of other
information from non-perturbative theoretical methods and experimental
input, one is forced to model $\rho_B^+(\omega',\mu)$ in this
region. To our knowledge there is no theorem stating that the original
LCDA $\phi_B^+(\omega,\mu)$ or even the dual $\rho_B^+(\omega',\mu)$
shall be positive definite, and therefore adopted models can differ
greatly. In the paper of which these proceedings report
\cite{Feldmann:2014ika} we have designed a recipe that allows for the
smooth merging of a given model in the low $\omega'$ region and the
perturbative findings of the previous section, while respecting the
moment constraints in (\ref{eq:momentconstraint}). This construction is
based on an expansion of the model in a set of basis functions with
apropriate characteristics. For details we refer the reader to the
original paper.

The first inverse moment of the LCDA -- and logarithmic modulations of
it -- are of particular interest to phenomenology. The factorized
amplitude of the exclusive $\bar B \to \gamma \ell \bar\nu$ in the low
$q^2$ region, for example, requires knowledge of the quantities
\cite{Beneke:2011nf,Braun:2012kp}
\begin{equation}
  \label{eq:15}
  \frac{\sigma_n(\mu)}{\lambda_B(\mu)} =\int_0^\infty \frac{d\omega}{\omega} \ln^n \left( \frac{\omega}{\mu} \right) \phi_B^+(\omega,\mu) \;,
\end{equation}
where $\lambda_B$ is of mass dimension 1 and $\sigma_n$ are numbers
with $\sigma_0 = 1$. Similarly we may define inverse moments in
dual space as
\begin{equation}
  \label{eq:16}
  L_n(\mu) = \int_0^\infty \frac{d\omega'}{\omega'} \ln^n \left( \frac{\hat\omega'}{\mu} \right) \rho_B^+(\omega',\mu) \;.
\end{equation}
It was shown that the first few of these dual moments are identical to
the original ones \cite{Bell:2013tfa}, namely
\begin{equation}
  \label{eq:17}
  L_0 = \frac{1}{\lambda_B} \;, \quad 
L_1 = \frac{\sigma_1}{\lambda_B}\;, \quad \hbox{and} \quad L_2 = \frac{\sigma_2}{\lambda_B}\;.
\end{equation}
For higher $n>2$ linear combinations appear. Since we already
know the integrand in (\ref{eq:16}) for $\hat\omega' \ge \mu$ we may
separate this region out and simply calculate it. The integral over
the remaining, non-perturbative part, $\hat\omega' \le \mu$, can be
rewritten by substituting $z = - \ln \frac{\hat\omega'}{\mu}$ to form 
\begin{equation}
  \label{eq:18}
  L_n^-(\mu) = \int_0^\infty dz\, (-z)^n \rho_B^+(\hat\mu e^{-z},\mu) \;.
\end{equation}
We may further expand the unknown function $\rho_B^+$ in terms of
Laguerre polynomials $\mathcal L_k(z)$, i.~e.
\begin{equation}
  \label{eq:19}
  \rho_B^+(\hat\mu e^{-z},\mu) = \sum\limits_{k=0}^\infty a_k(\mu)
  e^{-z} \mathcal L_k(z) \;.
\end{equation}
The advantage of this decomposition is that we may be able to fit
the first few coefficients $a_k(\mu)$ from precise experimental data
\cite{Beneke:2011nf,Braun:2012kp}. The above use of Laguerre
polynomials allows us to relate
\begin{eqnarray}
  \label{eq:20}
  L_0^-(\mu) &=& a_0(\mu) \;, \nnb \\
  L_1^-(\mu) &=& a_1(\mu) - a_0(\mu) \;, \nnb \\
  L_2^-(\mu) &=& 2a_2(\mu) - 4 a_1(\mu)+2a_0(\mu) \;, \nnb \\
 &\ldots &
\end{eqnarray}
In conclusion we find that at this point in time the most promissing
way to gain further insight into the non-perturbative part of the LCDA
is to increase the precision of experimental data from exclusive $B$
decays. Whether theoretical methods like sum rules and lattice QCD can
lead to more information on the LCDA is an interesting question.


\section{Conclusions.}
\nin 
In this talk we have summarized some aspects on the recently
found dual LCDA of the $B$ meson \cite{Feldmann:2014ika}. The biggest
advantage of this description results from the fact that it
renormalizes locally, i.~e.~does not mix different regions in its
argument $\omega'$ under RG evolution. The region of large $\omega'$
is determined by way of a short-distance operator product expansion of
moments over the original LCDA, and thus perturbatively calculable. We
have demonstrated this by calculating $\rho_B^+(\omega',\mu)$ in this
regime to first-order QCD corrections and first-order power
corrections. We advocate the use of the dual LCDA in factorization
theorems as it simplifies the RG analysis of the factorized amplitude
greatly. 

We paid particular attention to the tail of the dual LCDA, $\omega'
\gg \mu$, and demonstrated that resummation of large logarithms of the
form $\ln \hat\omega'/\mu$ renders the solution valid at any (and even
unphysically large) renormalization scale. Wherelse other formalisms
break down where the RG-evolution function $g$ assumes integer values,
our representation does not. 

We might be preempt in the notion that an impression exists, that an
increased precision in power corrections to the moments over the
original LCDA (\ref{eq:momentconstraint}) leads ultimately to a prediction of the
phenomenologically important quantity $\lambda_B$, but we stress that
this is not so. It would only lead to a more
precise determination of the dual LCDA in the large $\omega'$ regime,
but not determine $\lambda_B$ and its logarithmic modulations. On an
intuitive level this statement can be justified by the fact that the large
$b$-quark mass has been eliminated in HQET at leading power; on
a technical level it is justified by a very slow point-wise
convergence in the low-$\omega'$ region when expanding around the
large-$\omega'$ behaviour. For more details we refer the reader to the
original paper.

A similar treatment concerning the shape function of inclusive $B$
decays is possible.


\section*{Acknowledgements}
\nin
I would like to thank the organizers of this conference for a very
pleasant stay in Montpellier. I am supported by the Deutsche
Forschungsgemeinschaft within the Research Unit FOR 1873.


\end{multicols}

\begin{thebibliography}{999}
\vspace*{-0.25cm}

\bibitem{Beneke:1999br}
  M.~Beneke, G.~Buchalla, M.~Neubert and C.~T.~Sachrajda,
  Phys.\ Rev.\ Lett.\  {\bf 83} (1999) 1914
  [hep-ph/9905312].
  
\bibitem{Grozin:1996pq}
  A.~G.~Grozin and M.~Neubert,
  Phys.\ Rev.\  D {\bf 55} (1997) 272
  [arXiv:hep-ph/9607366].
  
\bibitem{Lee:2005gza}
  S.~J.~Lee, M.~Neubert,
  Phys.\ Rev.\ D {\bf 72} (2005) 094028
  [hep-ph/0509350].
 
\bibitem{Lange:2003ff}
  B.~O.~Lange, M.~Neubert,
  Phys.\ Rev.\ Lett.\  {\bf 91} (2003) 102001
  [hep-ph/0303082].

\bibitem{Bell:2013tfa}
  G.~Bell, Th.~Feldmann, Y.-M.~Wang and M.~W.~Y.~Yip,
  JHEP {\bf 1311} (2013) 191
  [arXiv:1308.6114 [hep-ph], arXiv:1308.6114].

  %
\bibitem{Efremov:1979qk}
  A.~V.~Efremov and A.~V.~Radyushkin,
  Phys.\ Lett.\ B {\bf 94} (1980) 245.

\bibitem{Lepage:1979zb}
  G.~P.~Lepage and S.~J.~Brodsky,
  Phys.\ Lett.\ B {\bf 87} (1979) 359;
  Phys.\ Rev.\ D {\bf 22} (1980) 2157.
  

\bibitem{Feldmann:2014ika} 
  T.~Feldmann, B.~O.~Lange and Y.~M.~Wang,
  Phys.\ Rev.\ D {\bf 89}, 114001 (2014)
  [arXiv:1404.1343 [hep-ph]].

\bibitem{Kawamura:2001jm}
  H.~Kawamura, J.~Kodaira, C.~-F.~Qiao, K.~Tanaka,
  Phys.\ Lett.\ B {\bf 523} (2001) 111
   [Erratum-ibid.\ B {\bf 536} (2002) 344]
  [hep-ph/0109181];
  H.~Kawamura and K.~Tanaka,
  Phys.\ Lett.\  B {\bf 673} (2009) 201
  [arXiv:0810.5628 [hep-ph]].
  
\bibitem{Beneke:2011nf}
  M.~Beneke and J.~Rohrwild,
  Eur.\ Phys.\ J.\ C {\bf 71} (2011) 1818
  [arXiv:1110.3228 [hep-ph]].
  
\bibitem{Braun:2012kp}
  V.~M.~Braun and A.~Khodjamirian,
  Phys.\ Lett.\ B {\bf 718} (2013) 1014
  [arXiv:1210.4453 [hep-ph]].
  
\end{thebibliography}
\end{document}